\documentstyle[12pt]{article}

\begin{document}

\begin{center}
\large{{\bf A MODEL TO EXPLAIN VARYING $\Lambda$, $G$ AND $\sigma^2$ SIMULTANEOUSLY}}
\end{center}

\medskip
\begin{center}
R. G. Vishwakarma\footnote{Email: rvishwa@mate.reduaz.mx}\\ 

\vspace{0.5cm}
\emph{Department of Mathematics\\
    Autonomous University of Zacatecas\\
  Zacatecas, ZAC C.P. 98060\\
                Mexico}
\end{center}

\medskip
\begin{abstract}\noindent
Models with varying cosmical parameters, which were earlier regarded constant, 
are getting attention. However, different models are usually invoked to 
explain the evolution of different parameters. We argue that whatever physical
process is responsible for the evolution of one parameter, should also be
responsible for the evolution of others. This means that the different 
parameters are coupled together somehow. Based on this guiding principle, we investigate a Bianchi type I model
with variable $\Lambda$ and $G$, in which $\Lambda$, $G$ and the shear parameter $\sigma^2$, all are coupled. It is interesting that the resulting model reduces to the FLRW model for large $t$ with $G$ approaching a constant.


\end{abstract}

\newpage
%\medskip
\noindent
{\bf 1. Introduction}

\noindent
In the past few years, evidence has mounted indicating that some \emph{constants},
which were earlier treated as true constants, are no longer constant in cosmology. The examples are $-$
Einstein's cosmological constant $\Lambda$, Newton's gravitational constant $G$, the fine structure constant, etc.
Different phenomenological models have been suggested to explain the evolutions
of different \emph{constants} (let us call them parameters). However, we believe
that there should be only one model to explain all these parameters if the underlying
theory is correct. Moreover, whatever physical
process is responsible for the evolution of one parameter, should also be
responsible for the evolution of others, implying that the different 
parameters are coupled together somehow.
It should, therefore, be the evolution of the universe itself
which should explain the dynamics of all the parameters.
In this paper, we investigate such a model from the Einstein field equations
which explains the variability of $\Lambda$, $G$ and the anisotropy parameter
$\sigma^2$ simultaneously. The cosmological consequences of the model are also
discussed.

Now we shall describe briefly the motivation for considering the different
parameters and their variations. The one which comes first in the list is 
undoubtedly the Einstein's cosmological parameter $\Lambda$, whose existence is favoured by the recent
supernovae (SNe) Ia observations \cite{sn} and which is also consistent with the recent anisotropy measurements of the
cosmic microwave background (CMB) made by the WMAP experiment \cite{cmb}. However, there is a fundamental
problem related with the existence of $\Lambda$, which has been extensively
discussed in the literature. It's value expected from the quantum field theory-
calculations is about 120 orders of magnitude higher than that estimated from the
observations. A phenomenological solution to this problem is suggested by considering 
$\Lambda$ as a function of time, 
so that it was large in the early universe and got reduced with the expansion of the universe \cite{vishwa}.

Variation of Newton's gravitational parameter $G$ was originally suggested
by Dirac on the basis of his large numbers hypothesis \cite{dirac}.
As $G$ couples geometry to matter, it is reasonable to consider $G=G(t)$
in an evolving universe when one considers $\Lambda=\Lambda(t)$. Many
extensions of general relativity with $G=G(t)$ have been made ever since
Dirac first considered the possibility of a variable $G$, though none
of these theories has gained wide acceptance. However a new approach, which
has been widely investigated in the past few years  \cite{wesson}, is 
appealing.
It assumes the conservation of the energy-momentum tensor which consequently 
renders $G$ and $\Lambda$ as coupled fields, similar to the case of $G$ in 
original
Brans-Dicke theory. This leaves Einstein's fields equations formally unchanged.
In this context, an approach is worth mentioning in which the scaling of
$G(t)$ and $\Lambda(t)$ arise from an underlying renormalization group
flow near an infrared attractive fixed point \cite{plb}. The resulting 
cosmology explains the high redshift SNe Ia and radio sources observations 
successfully \cite{reuter04}. It also describes the Planck era reliably and 
provides a resolution to the horizon and flatness problems of the 
standard cosmology without any unnatural fine tuning of the parameters
\cite{prd}. 
Gravitational theories with variable $G$ have also been discussed in the
context of induced gravity model where $G$ is generated by means of a 
non-vanishing vacuum expectation value of a scalar field \cite{zee}.
Recently a constraint on the variation of $G$ has been obtained by using WMAP 
and the big bang nucleosynthesis observations \cite{dotG}, which comes out as 
$-3 \times 10^{-13}$ yr$^{-1} < (\dot G/G)_{\rm today} < 4 \times 10^{-13}$ yr$^{-1}$.

Another important quantity which is supposed to be damped out in the course of 
cosmic evolution is the anisotropy of the cosmic expansion. It is believed 
that 
the early universe was characterized by a highly irregular expansion mechanism
which isotropized later \cite{misner}.
The level of anisotropy left out by the era of decoupling is only about
$10^{-5}$, as is revealed by the CMB observations.
It could be that whatever mechanism diminished $\Lambda$ to its present value, 
could have also rendered the early highly anisotropic universe to the present
smoothed out picture. This will be our guiding principle in investigating the
model. 

We shall keep
ourselves limited to Einstein's field equations and to the parameters which
appear explicitly therein. 
It would be worthwhile to mention that models with varying speed of light
are recently being promoted. These are supported by the claims, based on the 
measurements of distant quasar absorption spectra, that the fine structure constant
may have been smaller in the past. However, the speed of light $c$ has a
complex character having six different
facets which come from many laws of physics that are \emph{a priori} 
disconnected from the notion of light itself \cite{ellis2}. If it is the causal speed of which these theories are talking about, then one should not consider a varying $c$ in general relativity unless
the structure of the spacetime metric is changed and reinterpreted. We consider 
$c=1$ throughout our calculations.

We consider the Bianchi type I metric, which is the simplest anisotropic generalization 
of the flat Robertson-Walker metric and allows for different expansion factors in three
orthogonal directions. In the comoving coordinates ($u^i=\delta^i_0$), the
metric can be written as
\begin{equation}
{\rm d} s^2=-{\rm d}t^2+X^2(t)~{\rm d}x^2+Y^2(t)~{\rm d}y^2+Z^2(t)~{\rm d}z^2.
\label{eq:metric}
\end{equation}
An average {\it expansion scale factor} can be defined by
$R(t)=(XYZ)^{1/3}$ implying that the Hubble parameter $H=\dot R/R$.

\vspace{0.5cm}\noindent
{\bf 2. Field Equations}

\noindent
We consider $G$ and $\Lambda$ as functions of the cosmic time $t$.
For the metric (\ref{eq:metric}), the Einstein field equations, with perfect fluid, read
\begin{equation}
\frac{\dot X\dot Y}{XY}+\frac{\dot Y\dot Z}{YZ}+\frac{\dot Z\dot X}{ZX}=8\pi G\rho+\Lambda\\\label{eq:bian1}
\end{equation}
\begin{equation}
\frac{\ddot X}{X}+\frac{\ddot Y}{Y}+\frac{\dot X\dot Y}{XY}=-8\pi G w\rho+\Lambda\label{eq:bian2}\\
\end{equation}
\begin{equation}
\frac{\ddot Y}{Y}+\frac{\ddot Z}{Z}+\frac{\dot Y\dot Z}{YZ}=-8\pi G w\rho+\Lambda\label{eq:bian3}\\
\end{equation}
%\newpage
\begin{equation}
\frac{\ddot Z}{Z}+\frac{\ddot X}{X}+\frac{\dot Z\dot X}{ZX}=-8\pi G w\rho+\Lambda.\label{eq:bian4}\\
\end{equation}
Here we have assumed, as usual, an equation of state $p=w\rho$, 
where $0\leq w\leq 1$ is a constant. 
The non-vanishing components of the shear tensor $\sigma_{ij}$, defined by 
$\sigma_{ij}=u_{i;j}+u_{j;i}-\frac{2}{3}~g_{ij}~u^k_{~;k}$, are obtained as
\begin{equation}
\sigma^1_1=\frac{4}{3}\frac{\dot X}{X} - \frac{2}{3}\left(\frac{\dot Y}{Y} 
+ \frac{\dot Z}{Z} \right),
\end{equation}

\begin{equation}
\sigma^2_2=\frac{4}{3}\frac{\dot Y}{Y} - \frac{2}{3}\left(\frac{\dot Z}{Z} 
+ \frac{\dot X}{X} \right),
\end{equation}

\begin{equation}
\sigma^3_3=\frac{4}{3}\frac{\dot Z}{Z} - \frac{2}{3}\left(\frac{\dot X}{X} 
+ \frac{\dot Y}{Y} \right).
\end{equation}
Thus the magnitude $\sigma^2\equiv \sigma_{ij}\sigma^{ij}/8$ is obtained as
\begin{equation}
\sigma^2=\frac{1}{3}\left[\frac{{\dot X}^2}{X^2}+\frac{{\dot Y}^2}{Y^2}+\frac{{\dot Z}^2}{Z^2} -\left(\frac{\dot X\dot Y}{XY}+\frac{\dot Y\dot Z}{YZ}+\frac{\dot Z\dot X}{ZX} \right) \right].\label{eq:sigsq}
\end{equation}
It can be shown\footnote{By subtracting (\ref{eq:bian3}) from 
 (\ref{eq:bian2}), and  (\ref{eq:bian4}) from  (\ref{eq:bian3}) and integrating the resulting equations, one 
can get $\frac{\dot X}{X}-\frac{\dot Y}{Y} \propto \frac{1}{XYZ}$,   
$\frac{\dot Y}{Y}-\frac{\dot Z}{Z} \propto \frac{1}{XYZ}$,
 $\frac{\dot Z}{Z}-\frac{\dot X}{X} \propto \frac{1}{XYZ}$. By squaring and adding these equations one gets
 $\sigma^2 \propto 1/(XYZ)^2$.}
 that $\sigma^2$ is proportional to $R^{-6}$, i.e., $\sigma=\alpha R^{-3}$, 
where $\alpha=\rm constant$. This implies that
\begin{equation}
\frac{\dot \sigma}{\sigma}=-\left(\frac{\dot X}{X}+\frac{\dot Y}{Y}+\frac{\dot Z}{Z}\right)=-3H.\label{eq:forSig}
\end{equation}
Equations (\ref{eq:bian1}) and (\ref{eq:sigsq}) allow to write the analogue of the Friedmann equation as
\begin{equation}
3H^2=8\pi G\rho+\sigma^2+\Lambda.\label{eq:forH}
\end{equation}
So far, there has been no effect
of the varying characters of $G$ and $\Lambda$ on the equations and they are formally the same as
those with constant $G$ and $\Lambda$. However, the generalized conservation equation is different from the ordinary one. This can be obtained either from the
Bianchi identities or
by using equations (\ref{eq:bian2}$-$\ref{eq:bian4}) in the differentiated form  of equation (\ref{eq:bian1}) and can be written, after doing some simple algebra, in the 
form  
\begin{equation}
8\pi G~ [~\dot \rho+3(1+w)H\rho~] +8\pi\rho~\dot G+\dot\Lambda = 0.\label{eq:bianchi}
\end{equation}
We assume, as is common in cosmology, that the conservation of energy-momentum tensor of matter holds ($T^{ij}_{;j}=0$) leading to  
\begin{equation}
\dot \rho+3(1+w)H\rho =0,\label{eq:cons}
\end{equation}
leaving $G$ and $\Lambda$ as some kind of coupled fields:
\begin{equation}
8\pi\rho~\dot G+\dot\Lambda=0.\label{eq:forLam}
\end{equation}
Equation (\ref{eq:cons} has a simple solution $\rho=C R^{-3(1+w)}$, 
where ~$C$ = constant $>0$. 
Equation  (\ref{eq:forLam}) can be integrated as
\begin{equation}
G(R)=G_0-\frac{1}{8\pi C} \left[\Lambda (R) R^{3(1+w)} -3(1+w)\int \Lambda (R) R^{(2+3w)} dR\right],\label{eq:forG}
\end{equation}
where $G_0$ is a constant of integration. Equations (\ref{eq:forSig}$-$\ref{eq:forG}) 
supply only 4 independent equations in 5 unknowns $\rho, R, G, \Lambda$
and $\sigma$. In search of one more equation, we do some algebra in the following. 

An elimination of $H$ between (\ref{eq:forH}) and (\ref{eq:cons}) gives
\begin{equation}
\frac{\dot \rho^2}{\rho^3}=3(1+w)^2 \left(8\pi G+\frac{\sigma^2}{\rho}+\frac{\Lambda}{\rho}\right).
\end{equation}
Differentiating this and using (\ref{eq:forSig}), (\ref{eq:cons}) and (\ref{eq:forLam}) therein, we obtain
\begin{equation}
2\frac{\ddot \rho}{\rho}-3\frac{\dot \rho^2}{\rho^2}=3(1+w)^2 \left[\left(\frac{1-w}{1+w}\right)\sigma^2-\Lambda\right],w\ne -1,\dot\rho\ne 0,\label{eq:main1}
\end{equation}
which is the central equation of our investigation whose solution
will supply the required ansatz.
Substituting (\ref{eq:cons}) in (\ref{eq:main1}), we obtain an equation for $R$ as
\begin{equation}
\frac{2}{1+w}\dot H+3H^2+\left(\frac{1-w}{1+w}\right)\sigma^2-\Lambda=0.\label{eq:main2}
\end{equation}

\vspace{0.5cm}\noindent
{\bf 3. Models}

\noindent
If the physical processes, responsible for reducing the early highly anisotropic universe to a smooth present universe, are also responsible for bringing down
the large value of $\Lambda$ to its small present value, the two parameters
$\sigma^2$ and $\Lambda$ must be related somehow. In view of this guiding
principle, the simplest solution of equation (\ref{eq:main1}) is
\begin{equation}
\Lambda=\left(\frac{1-w}{1+w}\right)\sigma^2,\label{eq:lamTr}
\end{equation}
together with
\begin{equation}
2\frac{\ddot \rho}{\rho}=3\frac{\dot\rho^2}{\rho^2}.
\end{equation}
Equation (\ref{eq:lamTr}), which is our required ansatz, indicates a linear
coupling between the cosmological constant and anisotropy. The parameters 
$G$ and
$\Lambda$ are already coupled through equation (\ref{eq:forG}). We find 
that the model in this case is described by
\begin{equation}
R=a~t^{2/3(1+w)}, ~a=\rm constant >0,
\end{equation}
\begin{equation}
\rho=\left[\frac{C}{a^{3(1+w)}}\right] t^{-2},
\end{equation}
\begin{equation}
\sigma=\left[\frac{\alpha}{a^3}\right] t^{-2/(1+w)},
\end{equation}
\begin{equation}
\Lambda=\left[\left(\frac{1-w}{1+w}\right)\frac{\alpha^2}{a^6}\right] t^{-4/(1+w)},
\end{equation}
\begin{equation}
G=G_0-\left[\frac{\alpha^2}{4\pi C(1+w)} a^{-3(1-w)}\right] t^{-2(1-w)/(1+w)}.
\end{equation}
The model has a constant deceleration
parameter $q=(1+3w)/2$ and evolves to isotropy as $t\rightarrow \infty$,
with $\Lambda\rightarrow 0$ and $G\rightarrow G_0$.
Thus for large $t$, the model approaches the flat FLRW model which is very
encouraging. 
It may be noted that though the current observations of SNe Ia and CMB favour
accelerating models ($q<0$), but they do not altogether rule out the decelerating
ones which are also consistent with these observations \cite{vishwaN}. 
One can even fit the models with zero  $\Lambda$ if one takes
into account the extinction of light by the metallic dust ejected from the supernovae 
explosions \cite{vishwaN}.

We note that for $t< 3\alpha^2/16\pi C G_0 a^2$,  $G$ becomes negative
unless $w=1$ (with $G_0>\alpha^2/8\pi C$). 
One can however choose the constants $\alpha$ and $a$ (which are arbitrary)  
appropriately so that $G$ remains positive in the range of validity of general relativity.
However, taken at the face value, the model predicts a repulsive gravity in the range 
$0\leq t < 3\alpha^2/16\pi C G_0 a^2$.
For $w=1$, the model reduces to $\Lambda=0, G=\rm constant$ and $\sigma\propto H=1/3t$.

\medskip
The model can be generalized very easily by generalizing the ansatz (\ref{eq:lamTr}) in the form:
\begin{equation}
\Lambda=\gamma\sigma^2, ~ ~ \gamma=\rm constant,\label{eq:lam2}
\end{equation}
which can allow a negative $\Lambda$ as well (until we have precise enough SN Ia data
to rule out certain models, we should keep all our options open). Now with the new 
ansatz (\ref{eq:lam2}), equation (\ref{eq:forG}) reduces to
\begin{equation}
G=G_0 - \frac{\gamma\alpha^2}{4\pi C(1-w)} R^{-3(1-w)}, ~ ~  w\ne 1.
\end{equation}
The model starts from a 
big bang (or a big bang-like state) with $G,~ \mid \Lambda\mid$ ~ and $\sigma^2$ all infinite and evolves to
isotropy with $G\rightarrow G_0$ and $\Lambda\rightarrow 0$ as $t\rightarrow \infty$.
The time-evolution of $R$ is given by
%\newpage
\begin{equation}
t+t_0=\int \left[\frac{8\pi CG_0}{3}R^{-(1+3w)}+\frac{\alpha^2}{3} \left(1-\frac{1+w}{1-w}\gamma \right)R^{-4} \right]^{-1/2} dR, ~ ~  w\ne 1,\label{eq:forT}
\end{equation}
where $t_0$ is a constant of integration.
It is hard to integrate r.h.s. of equation (\ref{eq:forT}) for a general $w$ ($\ne 1$) unless 
$\gamma=(1-w)/(1+w)$ (which has already been investigated); 
or $G_0=0$ i.e., $\gamma<0$. If $G_0=0$, equation (\ref{eq:forT}) reduces to 
\begin{equation}
R=\alpha^{1/3} \left[3\left(1-\frac{1+w}{1-w}\gamma\right)\right]^{1/6} t^{1/3}.
\end{equation}
When $G_0\ne 0$, $\Lambda$ can assume both$-$ negative as well as positive values. 
In the case of a positive $\Lambda$, similar kind of argument, as above, can be given
when $G$ becomes negative in the beginning of the universe. For a positive $G_0$, 
equation (\ref{eq:forT}) can be integrated in different phases of evolution, as is shown in the following.

\vspace{0.7cm}
\noindent
{\bf $w=1/3:$}
\begin{equation}
t=\frac{1}{2\ell}\left[R\sqrt{\ell R^2+m}-\frac{m}{\sqrt \ell}\sinh^{-1} \left(\sqrt{\frac{\ell}{m}} R\right)\right],
\end{equation}

\vspace{0.7cm}\noindent
{\bf $w=0:$}

\begin{equation}
R=\left[\frac{9}{4}\ell(t+t_0)^2 -\frac{n}{\ell}\right]^{1/3},
\end{equation}
where $\ell=8\pi C G_0/3, \ m=(1-2\gamma)\alpha^2/3$,
$n=(1-\gamma)\alpha^2/3,\ t_0=2\sqrt{n}/3\ell$ 
and $\gamma < 1/2$.

\vspace{0.5cm}\noindent
{\bf 5. Conclusion}

\noindent
Einstein's field equations with time-dependent $G$ and $\Lambda$ have been
considered in the context of Bianchi type-I spacetime in such a way which
conserves the energy-momentum tensor of matter. 
We assume that the physical processes responsible for the evolution of one parameter, should also be
responsible for the evolutions of others. This means that the different 
parameters are coupled. In this view, the field equations
give a trivial ansatz implying a coupling between $\sigma^2$ (shear), $\Lambda$ and $G$.
The resulting model, for the baryonic matter, approaches the standard FLRW model
in the later epochs, with $G$ approaching a constant value. However, the earlier 
phases of the model are altogether different from that in the standard cosmology.
For stiff matter, the model reduces to the standard Bianchi type-I model with
$q=2, G=\rm constant, \Lambda=0, \rho\sim t^{-2}$ and $\sigma\sim H\sim t^{-1}$.

\vspace{0.5cm}\noindent
{\bf Acknowledgement}

\noindent
The author thanks the Abdus Salam ICTP for sending the necessary literature 
whenever required under the associateship programme.

\vspace{0.5cm}
%\newpage

\end{document}